\documentclass[prl,aps,twocolumn,superscriptaddress,floats,showpacs,amsmath,amssymb]{revtex4}

\usepackage{graphicx}
\usepackage{units}

\def \be{\begin{equation}}
\def \ee{\end{equation}}

\newcommand{\saclay}{%
           Nanoelectronics group, Service de Physique de l'Etat Condens{\'e},
           CEA Saclay F-91191 Gif-sur-Yvette Cedex, France}
\newcommand{\grenoble}{%
           INAC/SPSMS, CEA Grenoble, 17 rue des Martyrs, 
           38054 Grenoble Cedex, France}
\newcommand{\karlsruhe}{%
           Institut f{\"u}r Nanotechnologie,
           Forschungszentrum Karlsruhe, D-76021 Karlsruhe, Germany}
\newcommand{\strasbourg}{%
           Institut de Physique et Chimie des Mat{\'e}riaux de Strasbourg,
           UMR 7504 (CNRS-ULP), F-67034 Strasbourg Cedex 2, France}

\begin{document}

\title{Persistent currents in one dimension: 
       the other side of Leggett's theorem}

\author{Xavier Waintal}
\affiliation{\saclay}
\author{Genevi{\`e}ve Fleury}
\affiliation{\saclay}
\author{Kyryl Kazymyrenko}
\affiliation{\saclay}
\author{Manuel Houzet}
\affiliation{\grenoble}
\author{Peter Schmitteckert}
\affiliation{\karlsruhe}
\author{Dietmar Weinmann}
\affiliation{\strasbourg}

\date{\today}

\begin{abstract}
We discuss the sign of the persistent current of $N$ electrons in
one dimensional rings. Using a topology argument, we establish
lower bounds for the free energy in the presence of arbitrary 
electron-electron interactions and external potentials. 
Those bounds are the counterparts of upper bounds derived by 
Leggett. Rings with odd (even) numbers of polarized electrons are 
always diamagnetic (paramagnetic). We show that unpolarized 
electrons with $N$ being a multiple of four exhibit either 
paramagnetic behavior or a superconductor-like
current-phase relation.
\end{abstract}


\maketitle

A persistent current~\cite{buttiker1983} flows at low temperatures in 
small conducting rings when they are threaded by a magnetic flux $\phi$. 
This current is a thermodynamic effect which is deeply connected to the 
presence of quantum coherence. Its magnitude can be expressed as 
$I=-\partial F/\partial\phi$ in terms of the free energy $F$. 
Persistent currents have been in the focus of an intensive
theoretical activity (see for 
example~\cite{ambegaokar1990,schmid1991,vonoppen1991,altshuler1991,montambaux1997}). 
Nevertheless, the understanding of the experimentally measured currents
in metallic~\cite{levy1990,chandrasekhar1991,jariwala2001,deblock2002} and 
semiconductor~\cite{mailly1993,reulet1995} rings remains incomplete. In 
particular, the observed currents in diffusive rings are an order of 
magnitude larger than the value obtained from theories neglecting 
electron-electron interactions.

The persistent current itself generates a magnetic field which is detected
in the experiments. 
Besides the magnitude of the current, the sign of this magnetic response 
is of particular interest. 
In the absence of interactions, mesoscopic sample-to-sample fluctuations 
between paramagnetic ($F(\phi)<F(0)$) and diamagnetic ($F(\phi)>F(0)$) 
behavior are expected. Taking into account repulsive (attractive) 
interactions leads to the prediction of a paramagnetic (diamagnetic) 
average response in ensembles of diffusive rings. However, 
only diamagnetic signals have so far been observed in experiments.

A very general theoretical result in this domain is the theorem by 
Leggett~\cite{leggett1991} which states that the 
sign of the zero-temperature persistent current for spinless 
fermions (fully polarized electrons) in one dimensional (1D) rings is given 
by the parity of the particle number. This result appears in the form of 
upper bounds for the ground state energy. Leggett's theorem holds
for arbitrary potential landscape and electron-electron interactions. 

In this letter, we perform two tasks. Firstly, we establish general
lower bounds of the free energy that can be seen as the counterparts of 
Leggett's upper bounds. Our result is valid at any temperature and 
can be generalized to include the spin degree of freedom. 
Secondly, we use the density matrix renormalization group (DMRG) 
method~\cite{white92} to study scenarios allowed by those 
bounds.

We start with the Hamiltonian 
\be\nonumber
H=\sum_{i=1}^{N} \frac{1}{2m} [p_i-eA(r_i)]^2 +V(r_i) +\sum_{i<j} U(r_i-r_j)
\ee
describing $N$ spinless electrons in a 1D ring of size $L$,
where $p_i$ is the momentum of electron $i$, $A(r)$ the vector potential 
corresponding to a magnetic flux $\phi$, $V(r)$ 
an external potential and $U(r)$ the electron-electron interaction.
A gauge transformation allows to replace the vector potential by 
flux-dependent boundary conditions 
\be\nonumber
\Psi(L,r_2,\dots,r_N)=e^{i\Phi}\Psi(0,r_2,\dots,r_N)
\ee
with $\Phi=2\pi\phi e/h$ for the antisymmetric wavefunction $\Psi$.  

Leggett's argument~\cite{leggett1991} is based on the ansatz 
\be\nonumber
\Psi_v(r_1,\dots,r_N)=e^{i\theta(r_1,\dots,r_N)}\Psi_0
\ee 
for the variational ground state $\Psi_v$ at $\Phi\ne 0$ in terms of 
the exact ground state wavefunction $\Psi_0$ at $\Phi=0$.  
The energy corresponding to this ansatz is 
\be\nonumber
E_v=E(\Phi=0)+\frac{N\hbar^2}{2m}
\int d^Nr |\partial_{r_1}\theta|^2 |\Psi_0|^2 \, ,
\ee
where $E(\Phi)$ is the exact ground state energy. The phase $\theta$ 
is a symmetric function of its arguments and should go from 0 to $\Phi$
as $r_1$ goes around the ring from $0$ to $L$. In this loop, $r_1$ crosses 
the coordinates of the $N-1$ other particles. At each crossing the 
sign of $\Psi_0$ changes. 
Since $\Psi_0$ must be back to its initial value at the end of the 
loop, for even $N$ there must be at least \textit{one more} sign change. 
The function $\theta$ can be chosen such that its gradient is concentrated 
around this additional point where $\Psi_0$ vanishes, yielding $E_v=E(0)$.
This variational energy provides an upper bound $E(\Phi)\le E(0)$
for the ground state energy and the system is paramagnetic. A similar 
argument leads to $E(\Phi)\le E(\pi)$ for odd $N$. 
In the sequel, we show that these upper bounds can be complemented by 
corresponding lower bounds.

\begin{figure}
\includegraphics[keepaspectratio,width=0.85\linewidth]{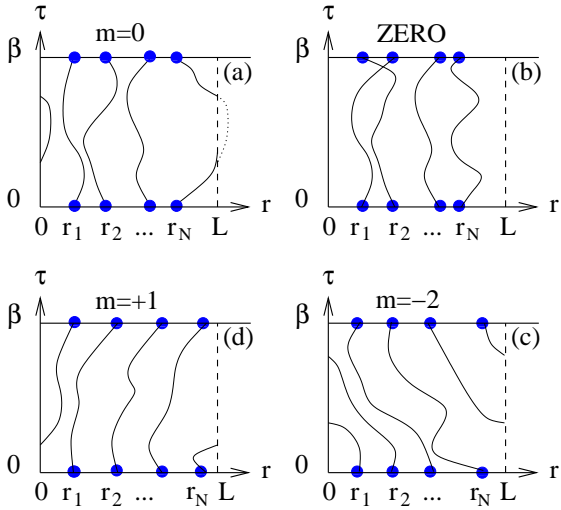}
\caption{\label{fig:paths} Four different paths $R(\tau)$
of the path integral \eqref{pathintegral}. The winding numbers $m$ are equal
to $0$ (a), $0$ (b), $-2$ (c) and $+1$ (d). Path (b) does not contribute
in 1D rings because two particle lines cross, which is forbidden by antisymmetry.}
\end{figure}
We consider the partition function $Z= {\rm Tr}_{\rm A} e^{-\beta H}$
where ${\rm Tr}_{\rm A}$ is the trace over antisymmetrized many-body states.
One can divide the inverse temperature $\beta$ in $K$ increments 
$\Delta\tau=\beta/K$ and insert $1=\int dR |R\rangle\langle R|$ at each 
step, where $|R\rangle=|r_1,r_2,\dots,r_N\rangle$ 
denotes the list of positions of all particles.
The result reads
\begin{eqnarray}
Z&=&\sum_P \sum_{R_1,\dots,R_K} |P| [e^{-\Delta\tau H}]_{R_1R_2}\
[e^{-\Delta\tau H}]_{R_2R_3}\dots
\nonumber \\ \nonumber
&& \dots [e^{-\Delta\tau H}]_{R_{K-1} R_K}\ [e^{-\Delta\tau H}]_{R_K P(R_1)}\, ,
\end{eqnarray}
where $P$ is a permutation of the $N$ particles and $|P|=\pm 1$
its signature. This path integral formula can be formally rewritten in 
continuous form as
\be\label{pathintegral}
Z=\sum_P |P| \int {\cal D }R(\tau) e^{-S[R(\tau)]}\, ,
\ee 
where $S[R(\tau)]$ is the action of the path $R(\tau)$ which satisfies 
$R(\beta)=P[R(0)]$. Examples for such paths 
are shown in Fig.~\ref{fig:paths}.
The effect of a magnetic flux on the path integral is 
a phase $\pm \Phi$ each time a particle crosses the boundary, yielding
\be\label{pathintegralflux}
Z(\Phi)=\sum_P |P| \int {\cal D }R(\tau)
e^{i\Phi m[R(\tau)]}
e^{-S[R(\tau)]}\, ,
\ee
where the winding number $m[R(\tau)]$ counts the total number of boundary 
crossings contained in $R(\tau)$ ($+1$ from left to right and $-1$ in 
the opposite direction). 
With $F=-\log Z/\beta$, Eq.~\eqref{pathintegralflux} leads to 
$\left.\partial^2 F/\partial\Phi^2\right|_{\Phi=0}=
\langle m^2[R]\rangle/\beta$
which is widely used to determine the superfluid fraction in bosonic systems,
in particular in the context of quantum Monte-Carlo 
simulations~\cite{pollock1987}. 
Anyway, Eq.~\eqref{pathintegralflux} can be expressed as
\be\nonumber
Z(\Phi)=\sum_{m=-\infty}^{+\infty} Z_m e^{i\Phi m}
\ee
with the partial partition functions $Z_m$ for a given $m$.

In 1D rings, most of the permutations $P$ do not contribute to the path 
integral since they would correspond to paths where two particles cross 
at some point. 
Those paths (see Fig.~\ref{fig:paths}b) have zero contribution due to 
antisymmetry. 
As a result, only paths with \textit{cyclic permutations} contribute to the 
path integral. There is only one single cyclic permutation $P_m$ per winding 
number $m$ (see Fig.~\ref{fig:paths}), whose signature determines 
the sign of the corresponding $Z_m$. Since the signature of a cyclic 
permutation of $N$ particles with winding number $m$ is given by 
$|P_m|=(-1)^{m (N-m)}$, we have
\be\nonumber
Z_m=(-1)^{m (N-m)} |Z_m|\, .
\ee

If $N$ is odd, then $m (N-m)$ is even and $Z_m$ is 
positive for all  $m$. Since $Z_m=Z_{-m}$, we have
\be\nonumber
Z(\Phi)=Z_0 +2 \sum_{m=1}^{\infty} Z_m \cos (m\Phi) \le
Z_0 +2 \sum_{m=1}^{\infty} |Z_m|  =  Z(0)
\ee
so that $F(\Phi)\ge F(0)$, and the system is diamagnetic.
When $N$ is even, $m(N-m)$ has the same parity as $m$
and $Z_m=(-1)^m |Z_m|$, leading to
\be\nonumber
Z(\Phi) \le 
Z_0 +2 \sum_{m=1}^{\infty} |Z_m|  =  Z(\pi) 
\ee 
such that $F(\Phi)\ge F(\pi)$.  
At zero temperature, the above inequalities together with  
those obtained by Leggett read   
\begin{eqnarray} 
\label{odd_p} 
E(0)\le E(\Phi)\le E(\pi) &\mbox{ for }&  N\ {\rm odd},\\ 
\label{even_p} 
E(\pi)\le E(\Phi)\le E(0) &\mbox{ for }&  N\ {\rm even}.
\end{eqnarray} 
 
The previous results can be partially extended to unpolarized 1D systems
with $N_\uparrow$ up and $N_\downarrow$ down electrons under a spin 
conserving Hamiltonian. 
Introducing winding numbers $m_\uparrow$ 
and $m_\downarrow$ for the two species, we have 
\be\nonumber 
Z(\Phi)=\sum_{m_\uparrow ,m_\downarrow} Z_{m_\uparrow ,m_\downarrow}  
 e^{i\Phi (m_\uparrow +m_\downarrow)} \, .
\ee 
Using similar arguments as above, we find
$Z_{m_\uparrow ,m_\downarrow}=(-1)^{|P_{m_\uparrow} |+|P_{m_\downarrow}|} 
|Z_{m_\uparrow ,m_\downarrow}|$
and conclude that 
\begin{eqnarray} 
\label{odd_np} 
F(0)\le F(\Phi) &\mbox{ for }& 
N_\uparrow \ {\rm and}\ N_\downarrow \ {\rm odd},\\ 
\label{even_np} 
F(\pi)\le F(\Phi) &\mbox{ for }& 
N_\uparrow \ {\rm and}\ N_\downarrow \ {\rm even}. 
\end{eqnarray} 
Those lower bounds are valid at arbitrary temperature. 
Eqs.~\eqref{odd_p} and \eqref{even_p} for the polarized case and 
\eqref{odd_np} and \eqref{even_np} for unpolarized electrons 
are the central results of this work. 
 
\begin{figure}
\includegraphics[keepaspectratio,width=\linewidth]{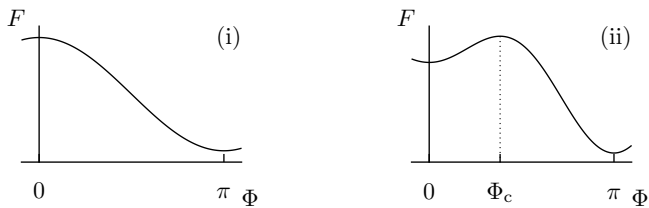}
\caption{\label{fig:scenario} Two possible flux-dependences of the 
free energy $F$ for $N=4n$ with even $N_\uparrow=N_\downarrow$, corresponding 
to paramagnetic (i) and diamagnetic (ii) behavior.}
\end{figure}
We now consider rings with a number of electrons
$N=N_\uparrow+N_\downarrow=4n$ being a multiple of four, and 
characterize the flux-dependence of the free energy using the 
curvatures 
\be\nonumber
C_{0(\pi)}=\partial^2 F/\partial\Phi^2|_{\Phi=0(\pi)}
\ee
at $\Phi=0$ and $\pi$.
In the unpolarized case ($N_\uparrow=N_\downarrow=2n$)
Eq.~\eqref{even_np} applies, the free energy assumes a 
minimum at $\Phi=\pi$, and $C_\pi$ is positive.  
The curvature at zero flux $C_0$ is not constrained such that the two 
scenarios illustrated in Fig.~\ref{fig:scenario} remain.  
(i) If $C_0$ is negative, the system is paramagnetic and lowers its energy 
in the presence of a magnetic flux.    
(ii) If $C_0>0$, the system is diamagnetic. In this case
the usual sinusoidal current-phase relation $F\propto-\cos\Phi$ is  
prohibited by $C_\pi>0$. There is a strong contribution of the second 
harmonic in $\Phi$ and it exists at least one flux value 
$0<\Phi_\mathrm{c}<\pi$ where the free energy is maximum and the persistent 
current vanishes. Without interaction, one finds case (i). Case (ii) is 
characteristic, for instance, of superconducting fluctuations induced by a 
weak attractive interaction (in a superconductor, 
$F\propto -\cos(2\Phi)$ and $\Phi_\mathrm{c}=\pi/2$). 

The overall conclusion is that a simple diamagnetic response is 
prohibited. The system must either be in a paramagnetic 
state or have a superconducting-like current-phase relation. 
 
In order to illustrate the behavior of the curvatures  
$C_0$ and $C_\pi$, we consider 1D tight-binding rings of $M$  
sites, described by the Hamiltonian 
\begin{eqnarray}\label{hubbard} 
H&=&\sum_\sigma\sum_{j=1}^{M} 
\left\{ -t \left( c^{\dagger}_{j+1,\sigma}c^{\phantom{\dagger}}_{j,\sigma}
        +\mbox{h.c.} \right) 
+V_{j,\sigma}n_{j,\sigma} 
\right\} 
\nonumber \\ 
&+&\sum_{j=1}^{M}\left(U_0n_{j,\uparrow}n_{j,\downarrow} 
                      +U_1n_{j+1}n_{j}\right) \, . 
\end{eqnarray} 
Here, $c^{\dagger}_{j,\sigma}$ creates an electron with spin  
$\sigma=\{\uparrow,\downarrow\}$ on site $j$, the random on-site  
energies $V_{j,\sigma}$ are drawn independently from the interval  
$[-W/2,W/2]$, and the kinetic energy scale is given by the hopping  
amplitude $t$.   
We include Hubbard on-site and nearest-neighbour interactions of strength  
$U_0$ and $U_1$, respectively. The density operators are defined as  
$n_{j,\sigma}=c^{\dagger}_{j,\sigma}c^{\phantom{\dagger}}_{j,\sigma}$ and  
$n_{j}=n_{j,\uparrow}+n_{j,\downarrow}$, and
the boundary condition $c_{M+1,\sigma}=e^{i\Phi}c_{1,\sigma}$ accounts for a  
magnetic flux threading the ring.  
 
We use the DMRG algorithm~\cite{white92} adapted to disordered 
systems~\cite{schmitteckert99} to 
calculate the ground state energies and the zero-temperature  
curvatures~\cite{schmitteckert04} for $\Phi=0$ and $\Phi=\pi$, fully 
taking into account the many-body correlations. For the largest systems 
750 states per block are kept in the DMRG iterations.    
All of the numerical results should, and do, satisfy the 
relations \eqref{odd_np}  and \eqref{even_np}. 

\begin{figure} 
\includegraphics[keepaspectratio,width=\linewidth]{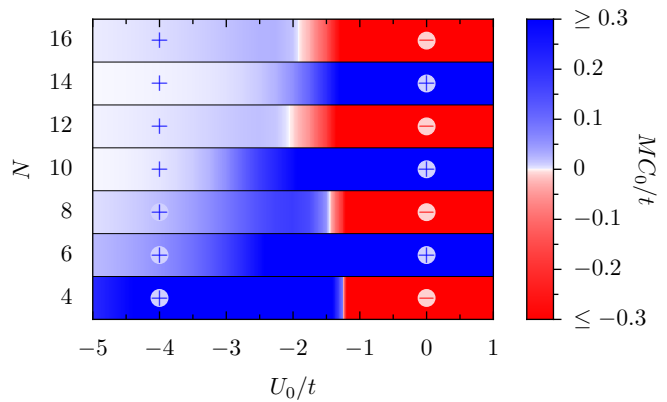} 
\caption{\label{fig:super} Curvatures $C_0$ at zero temperature for 
disordered quarter-filled Hubbard rings of different sizes $M=2N$, as 
a function of $U_0$, at $U_1=0$ and $N_\uparrow=N_\downarrow$, for 
one disorder realization with $W=t$.} 
\end{figure} 
In Fig.~\ref{fig:super}, we show the effect of an attractive interaction 
on the curvature $C_0$, for single realizations of disordered 
quarter-filled Hubbard rings of sizes up to $M=2N=32$ sites with 
$W=t$ and $U_1=0$. 
For even $N_\uparrow=N_\downarrow$ ($N=4n$), an attractive 
interaction reverses the sign of $C_0$. Hence the interactions induce a 
transition from scenario (i) to (ii) of Fig.~\ref{fig:scenario} and drive 
the system from paramagnetic towards ``superconducting''.
In contrast, the sign of $C_0$ for odd $N_\uparrow=N_\downarrow$ $(N=4n+2)$   
remains positive for all values of $U_0$, as dictated by \eqref{odd_np}. 
  
\begin{figure}
\includegraphics[keepaspectratio,width=0.9\linewidth]{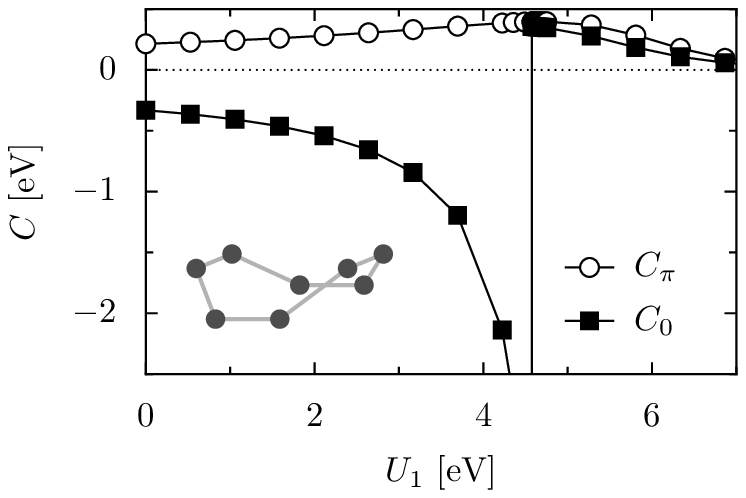} 
\caption{\label{fig:cyclo} Curvatures $C_0$ and $C_\pi$ as a function of $U_1$
for an unpolarized clean ring with $M=8$ at zero temperature, for the 
parameters of cyclooctatetraene 
($N=8$, $t=\unit[2.64]{eV}$, $U_0=\unit[8.9]{eV}$). 
Inset: Sketch of the C atom configuration in cyclooctatetraene.} 
\end{figure}

A molecular realization of the $N=4n$ case is  
cyclooctatetraene ($\mathrm{C}_8\mathrm{H}_8$, see the inset of  
Fig.~\ref{fig:cyclo}), which consists of a ring of eight carbon atoms with 
eight $\pi$ electrons. 
In Fig.~\ref{fig:cyclo}, we plot the curvatures $C_0$ and $C_\pi$ as 
a function of the nearest-neighbour interaction strength $U_1$ for the 
model Hamiltonian \eqref{hubbard} with $M=8$ and $W=0$, 
using the parameters $t=\unit[2.64]{eV}$ and $U_0=\unit[8.9]{eV}$ given in 
Ref.~\cite{himmerich2006} for cyclooctatetraene.  
Depending on the strength of $U_1$ (and the neglected 
longer range parts of the interaction), the system can undergo a 
transition~\cite{jeckelmann2002}  
from a paramagnetic spin-density-wave ($U_1\le \unit[4.6]{eV}$ and $C_0<0$) 
to a diamagnetic charge-density-wave ($U_1\ge \unit[4.6]{eV}$ and $C_0>0$) 
groundstate.  

The presence of paramagnetism or orbital ferromagnetism~\cite{buzdin84} in 
cyclooctatetraene is still a matter of 
debate~\cite{havenith2006,himmerich2006}. 
A ferromagnetic instability can occur in small paramagnetic rings provided
their inductance is large enough. Then, a magnetic field fluctuation
generates a current which reinforces the magnetic field. 
For small flux, the persistent current is $I\approx C_0 e \Phi/\hbar$ and 
the magnetic flux generated by this current is  
$\Phi\approx \mu_0 e L I/\hbar$ ($L\approx \unit[1]{nm}$ is the typical 
circumference of the molecule and $\mu_0L$ is its typical inductance). 
The instability occurs when the magnetic field generated   
current amplifies field fluctuations, i.e. when $C_0<0$ and
$X=\mu_0 |C_0| e^2 L /\hbar^2>1$. In our model, the most pronounced 
negative curvature $C_0\approx \unit[-5]{eV}$ occurs in the 
spin-density-wave regime close to the transition at 
$U_1\approx \unit[4.6]{eV}$. 
We therefore have $X\lesssim 3\times 10^{-3}$, far from the ferromagnetic 
instability. Moreover, Jahn-Teller distortions in cyclotetraene have been 
predicted~\cite{havenith2006} which would reduce the curvatures. 

An alternative to small molecules are rings made in semiconductor 
heterostructures. 
Important progress has been realized in the fabrication of such systems and  
1D rings with a single conduction channel could be produced.  
Estimates show that one still has $X\ll 1$. Therefore, no orbital 
ferromagnetism is to be expected, reminiscent with the result for large 1D 
Luttinger liquid rings~\cite{loss1993}.

However, as illustrated by the
results shown in Fig.~\ref{fig:super}, we predict that  
changing the number of electrons by two (with a back gate for instance) 
leads to a change of either the sign of the magnetic response (paramagnetism) 
or the periodicity of the response to a magnetic field 
(``superconducting like''). This could allow to detect small 
attractive interactions in those systems, whose presence is suggested by  
the fact that so far only diamagnetic responses have been observed in  
multichannel rings. 
 
In conclusion, we have established general relations for the sign of 
persistent currents in 1D systems that can be considered as the 
counterpart of Leggett's theorem for spinless fermions.  
Our theorem includes spin and is valid at arbitrary temperature. 
For spinless fermions, our relations imply that 
interactions and disorder cannot affect the sign of the persistent current. 
In particular, the possibility discussed by Leggett of 
having maxima of the energy at both, integer and half integer flux quantum, 
corresponding to a paramagnetic signal in an assembly of rings, is  
ruled out by \eqref{odd_p} and \eqref{even_p}. 
For electrons with spin, when Leggett's theorem does not apply, only 
the lower bounds \eqref{odd_np} and \eqref{even_np} for the free energy 
remain.  
We showed that this allows for a ``superconducting-like'' current-phase  
relation when attractive electron-electron interactions change the sign 
of the unconstrained curvature. Our topological argument provides a rigorous  
justification for the phenomenological H\"uckel rule which states  
that cyclic molecules with $4n+2$ electrons like benzene are aromatic 
while those with $4n$ electrons are not.  

\bibliography{persistent1D} 
\end{document}